\documentclass[12pt]{article}
\usepackage{a4}
\usepackage{amsfonts}
\usepackage{amsmath,amssymb}

\makeatother

\def\un#1{\relax\ifmmode\@@underline#1\else
        $\@@underline{\hbox{#1}}$\relax\fi}

\let\du=\du                     

\def\a{\alpha}
\def\b{\beta}

\def\d{\delta}

\def\f{\phi}
\def\g{\gamma}

\def\j{\psi}

\def\m{\mu}
\def\n{\nu}

\def\s{\sigma}

\def\x{\xi}

\def\F{\Phi}

\def\L{\Lambda}
\def\O{\Omega}

\def\ce{{\cal E}}

\def\cg{{\cal G}}

\def\car{{\cal R}}

\def\cv{{\cal V}}
\def\cw{{\cal W}}

\def\cz{{\cal Z}}

\def\bo{{\raise-.3ex\hbox{\large$\Box$}}}               
\def\pa{\partial}                                       
\def\de{\nabla}                                         
\def\TH{{\raise.2ex\hbox{$\displaystyle \bigodot$}\mskip-4.7mu \llap H \;}}
\def\face{{\raise.2ex\hbox{$\displaystyle \bigodot$}\mskip-2.2mu \llap {$\ddot
        \smile$}}}                                      

\def\VEV#1{\left\langle #1\right\rangle}        
\def\abs#1{\left| #1\right|}                    
\def\leftrightarrowfill{$\mathsurround=0pt \mathord\leftarrow \mkern-6mu
        \cleaders\hbox{$\mkern-2mu \mathord- \mkern-2mu$}\hfill
        \mkern-6mu \mathord\rightarrow$}
\def\dvec#1{\vbox{\ialign{##\crcr
        \leftrightarrowfill\crcr\noalign{\kern-1pt\nointerlineskip}
        $\hfil\displaystyle{#1}\hfil$\crcr}}}           
\def\dt#1{{\buildrel {\hbox{\LARGE .}} \over {#1}}}     

\def\frac#1#2{{\textstyle{#1\over\vphantom2\smash{\raise.20ex
        \hbox{$\scriptstyle{#2}$}}}}}                   
\def\sfrac#1#2{{\vphantom1\smash{\lower.5ex\hbox{\small$#1$}}\over
        \vphantom1\smash{\raise.4ex\hbox{\small$#2$}}}} 
\def\bfrac#1#2{{\vphantom1\smash{\lower.5ex\hbox{$#1$}}\over
        \vphantom1\smash{\raise.3ex\hbox{$#2$}}}}       
\def\afrac#1#2{{\vphantom1\smash{\lower.5ex\hbox{$#1$}}\over#2}}    

\def\[{\lfloor{\hskip 0.35pt}\!\!\!\lceil}
\def\]{\rfloor{\hskip 0.35pt}\!\!\!\rceil}
\def\Lag{{\cal L}}
\def\du#1#2{_{#1}{}^{#2}}

\def\fracm#1#2{\hbox{\large{${\frac{{#1}}{{#2}}}$}}}
\def\ha{{\fracmm12}}

\def\un{\underline}
\def\fracmm#1#2{{{#1}\over{#2}}}

\def\low#1{{\raise -3pt\hbox{${\hskip 0.75pt}\!_{#1}$}}}

\def\Dot#1{\buildrel{_{_{\hskip 0.01in}\bullet}}\over{#1}}
\def\dt#1{\Dot{#1}}

\newskip\humongous \humongous=0pt plus 1000pt minus 1000pt
\def\caja{\mathsurround=0pt}
\def\eqalign#1{\,\vcenter{\openup2\jot \caja
        \ialign{\strut \hfil$\displaystyle{##}$&$
        \displaystyle{{}##}$\hfil\crcr#1\crcr}}\,}
\newif\ifdtup

\newcommand{\be}{\begin{equation}}
\newcommand{\ee}{\end{equation}}
\newcommand{\nbe}{\begin{equation*}}
\newcommand{\nee}{\end{equation*}}

\newcommand{\lb}{\label}

\begin{document}

\thispagestyle{empty}

{\hbox to\hsize{
\vbox{\noindent March 2009 \hfill version 2 }}}

\noindent
\vskip2.0cm
\begin{center}

{\large\bf SCALAR POTENTIAL IN $F({\cal R})$ SUPERGRAVITY~\footnote{Supported 
in part by the Japanese Society for Promotion of Science (JSPS)}}
\vglue.3in

Sergei V. Ketov
\vglue.1in

{\it Department of Physics, Tokyo Metropolitan University, Japan}
\vglue.1in
ketov@phys.metro-u.ac.jp
\end{center}

\vglue.3in

\begin{center}
{\Large\bf Abstract}
\end{center}
\vglue.1in

\noindent We derive a scalar potential in the recently proposed N=1 
supersymmetric generalization of f(R) gravity in four space-time dimensions.  
Any such higher-derivative supergravity is classically equivalent to the 
standard N=1 supergravity coupled to a chiral (matter) superfield, via a 
Legendre-Weyl transform in superspace. The Kaehler potential, the 
superpotential and the scalar potential of that theory are all governed by a 
single holomorphic function. We also find the conditions for the vanishing 
cosmological constant and spontanenous supersymmetry breaking, without 
fine-tuning, which define a no-scale F(R) supergravity. The F(R) supergravities
are suitable for physical applications in the inflationary cosmology based on 
supergravity and superstrings.

\newpage

\section{Introduction}

Superstring theory naturally includes gauge field theories and gravity, it 
offers a consistent perturbation theory for quantum gravity, and it has promise
of unifying the Standard Model of elementary particles with gravity. An 
experimental test of string theory may be possible by future precision data 
{}from cosmological observations. However, it requires, in the first place, a 
reliable theoretical derivation of inflation from string theory  
\cite{pert,kall,iihk}.

String theory needs local supersymmetry for its consistency, whereas the most
direct connection to the observational cosmology is provided by the effective
N=1 supergravity in four space-time dimensions, after a flux compactification 
and moduli stabilization of the ten-dimensional superstring theory (or 
M-theory in eleven dimensions) \cite{fluxes}.

In our recent paper \cite{gket} we proposed the geometrical origin of an
inflaton and quintessence, as described by a dynamically generated complex 
scalar field in a certain higher-derivative N=1 supergravity theory. Our 
modified supergravity \cite{gket} can be seen as the N=1 locally 
supersymmetric extension of the $f(R)$ gravity theories that attracted 
considerable interest in recent years due to their natural ability to 
describe the current universe acceleration and unify it with the early universe
 inflation \cite{sot}. 

The supersymmetric extension of the $f(R)$ gravity theories we proposed in
ref.~\cite{gket} is non-trivial because, despite of the apparent presence 
of the higher derivatives of arbitrary order in the Lagrangian, there are no 
ghosts, potential instabilities can be avoided, and the auxiliary freedom 
\cite{gat} is preserved. Perhaps, most importantly, the modified supergravity 
action \cite{gket} is classically equivalent to the {\it standard} N=1 
supergravity minimally coupled to a chiral `matter' superfield whose K\"ahler
potential and superpotential are dictated by a single holomorphic function 
(see Secs. 2 and 3 below).

We speculated in ref.~\cite{gket} that our $F(\cal R)$ supergravity may be 
dynamically generated by some non-perturbative quantum superstring corrections,
whereas the complex scalar field component of the physical chiral scalar 
superfield (in the equivalent scalar-tensor supergravity) may be identified 
with a dilaton-axion pair.

In this Letter we would like to investigate some model-independent 
phenomenological prospects of our proposal \cite{gket} by deriving a scalar 
potential in our modified supergravity. We get the conditions for the vanishing
 cosmological constant together with spontaneous supersymmetry breaking, 
without fine tuning. Those no-scale supergravities are the starting point of 
almost any derivation of inflation from supergravity and string theory --- 
see e.g., ref.~\cite{kall} and references therein. 

In Sec.~2 we give our notation and setup. In Sec.3 we recall (and generalize) 
the derivation \cite{gket} of the classical equivalence via a Legendre-Weyl 
transform in curved superspace, and add a discussion of the super-K\"ahler-Weyl
 invariance. In Sec.~4 we introduce the no-scale modified  supergravity. Sec.~4
 is our conclusion. 

\section{Notation and setup}

A concise and manifestly supersymmetric description of supergravity is given
by superspace \cite{sspace}. In this section we provide a few equations for
notational purposes only.~\footnote{We use the natural units 
$c=\hbar=\kappa=1$.}

The chiral superspace density (in the supersymmetric gauge-fixed form) is
\be \lb{den}
\ce(x,\theta) = e(x) \left[ 1 -2i\theta\s_a\bar{\j}^a(x) +
\theta^2 B(x)\right]~, \ee
where $e=\sqrt{-\det g_{\m\n}}$, $g_{\m\n}$ is a spacetime metric, 
$\j^a_{\a}=e^a_{\m}\j^{\m}_{\a}$ is a chiral gravitino, $B=S-iP$ is the 
complex scalar auxiliary field. We use the lower case middle greek letters 
$\m,\n,\ldots=0,1,2,3$ for curved spacetime vector indices, the lower case 
early latin letters $a,b,\ldots=0,1,2,3$ for flat (target) space vector 
indices, and the lower case early greek letters $\a,\b,\ldots=1,2$ for chiral
 spinor indices.

The solution of the superspace Bianchi identitiies and the constraints defining
the $N=1$ Poincar\'e-type minimal supergravity results in only three relevant 
superfields $\car$, $\cg_a$ and $\cw_{\a\b\g}$ (as parts of the supertorsion), 
subject to the off-shell relations \cite{sspace}
\be \lb{bi1}
 \cg_a=\bar{\cg}_a~,\qquad \cw_{\a\b\g}=\cw_{(\a\b\g)}~,\qquad
\bar{\de}_{\dt{\a}}\car=\bar{\de}_{\dt{\a}}\cw_{\a\b\g}=0~,\ee
and
\be \lb{bi2}
 \bar{\de}^{\dt{\a}}\cg_{\a\dt{\a}}=\de_{\a}\car~,\qquad
\de^{\g}\cw_{\a\b\g}=\frac{i}{2}\de\du{\a}{\dt{\a}}\cg_{\b\dt{\a}}+
\frac{i}{2}\de\du{\b}{\dt{\a}}\cg_{\a\dt{\a}}~~,\ee
where $(\de\low{\a},\bar{\de}_{\dt{\a}}.\de_{\a\dt{\a}})$ represent the curved 
superspace $N=1$ supercovariant derivatives, and bars denote complex 
conjugation.

The covariantly chiral complex scalar superfield $\car$ has the scalar 
curvature $R$ as the coefficient at its $\theta^2$ term, the real vector 
superfield $\cg_{\a\dt{\a}}$ has the traceless Ricci tensor, 
$R_{\m\n}+R_{\n\m}-\frac{1}{2}g_{\m\n}R$, as the coefficient at its 
$\theta\s^a\bar{\theta}$ term, whereas the covariantly chiral, complex, 
totally symmetric, fermionic superfield $\cw_{\a\b\g}$ has the Weyl tensor 
$W_{\a\b\g\d}$ as the coefficient at its linear $\theta^{\d}$-dependent term. 

A generic supergravity Lagrangian (e.g. representing the supergravitational 
part of the superstring effective action) is
\be \lb{genc}
\Lag = \Lag(\car,\cg,\cw,\ldots) \ee
where the dots stand for arbitrary covariant derivatives of the supergravity 
superfields. We would like to concentrate on the particular 
sector of the theory (\ref{genc}), by ignoring the tensor superfields 
 $\cw_{\a\b\g}$ and $\cg_{\a\dt{\a}}$, and the derivatives of the 
scalar superfield $\car$. Thus the effective higher-derivative supergravity 
acton we proposed \cite{gket} in given by 
\be
\lb{action}
 S_F = \int d^4xd^2\theta\,\ce F(\car) + {\rm H.c.}
\ee
with some holomorphic function $F(\car)$ presumably generated by strings. 
Besides manifest local $N=1$ supersymmetry, the action (\ref{action}) also
possess the auxiliary freedom \cite{gat}, since the auxiliary field $B$ 
does not propagate. It distinguishes the action (\ref{action}) from other 
possible truncations of eq.~(\ref{genc}). In addition, the action 
(\ref{action}) gives rise to a spacetime torsion.

\section{Super-Legendre-Weyl-K\"ahler transform}

The superfield action (\ref{action}) is classically equivalent to another 
action
\be \lb{lmult}
 S_V = \int d^4x d^2\theta\,\ce \left[ \cz\car -V(\cz)\right] + {\rm H.c.}
\ee
where we have introduced the covariantly chiral superfield $\cz$ as a 
 Lagrange multiplier. Varying the action (\ref{lmult}) with respect to 
$\cz$  gives back the original action  (\ref{action}) provided that
\be \lb{lt1} 
F(\car) =\car\cz(\car)-V(\cz(\car)) \ee
where the function $\cz(\car)$ is defined by inverting the function
\be \lb{lt2}
\car =V'(\cz) \ee

Equations (\ref{lt1}) and  (\ref{lt2}) define the superfield Legendre 
transform. They imply further relations,
\be \lb{lt3}
F'(\car)=Z(\car)\qquad {\rm and}\qquad F''(\car)=Z'(\car)
=\fracmm{1}{V''(\cz(\car))}  \ee
where $V''=d^2V/d\cz^2$. The second formula (\ref{lt3}) is the duality relation
 between the supergravitational function $F$ and the chiral superpotential $V$.

A super-Weyl transform of the superfeld acton (\ref{lmult}) can also be done
entirely in superspace, i.e. with manifest local N=1 supersymmetry. In terms
of components, the super-Weyl transform amounts to a Weyl transform, a chiral 
rotation and a (superconformal) $S$-supersymmetry transformation \cite{howe}.

The chiral density superfield $\ce$ is a chiral compensator of the 
super-Weyl transformations
\be \lb{swt}
\ce \to e^{3\F} \ce~, \ee
whose parameter $\F$ is an arbitrary covariantly chiral superfield,
$\bar{\de}_{\dt{\a}}\F=0$. Under the transformation (\ref{swt}) the 
covariantly chiral superfield $\car$ transforms as 
\be \lb{rwlaw}
\car \to e^{-2\F}\left( \car - \fracm{1}{4}\bar{\nabla}^2\right)
e^{\bar{\F}}
\ee

The super-Weyl chiral superfield parameter $\F$ can be traded for the chiral
Lagrange multiplier $\cz$ by using a generic gauge condition~\footnote{In 
ref.~\cite{gket} we used the particular gauge $\x\F =\ln\cz$ with a number 
$\x$.} 
\be \lb{ch2} \cz=\cz(\F) \ee
where $\cz(\F)$ is an arbitrary (holomorphic) function of $\F$. Then the 
super-Weyl transform of the acton (\ref{lmult}) results in the classically 
equivalent action   
\be \lb{chimat2}
S_{\F} =  \int d^4x d^4\theta\, E^{-1} e^{\F+\bar{\F}}
\left[ \cz(\F) +\bar{\cz}(\bar{\F}) \right] 
+\left[ - \int d^4x d^2\theta\, \ce e^{3\F}V(\cz(\F)) +{\rm H.c.} 
\right]~,
\ee
where we have introduced the full supergravity supervielbein  $E^{-1}$ 
\cite{sspace}.

Equation (\ref{chimat2}) has the standard form of the action of a chiral matter
superfield coupled to supergravity \cite{sspace},
\be \lb{stand}
S[\F,\bar{\F}]= \int d^4x d^4\theta\, E^{-1} \O(\F,\bar{\F}) 
+\left[ \int d^4x d^2\theta\, \ce P(\F) +{\rm H.c.} \right]~,
\ee  
in terms of a `K\"ahler' potential $\O(\F,\bar{\F})$ and a chiral 
superpotential $P(\F)$. In our case (\ref{chimat2}) we find
\be \lb{spots}
\eqalign{
\O(\F,\bar{\F}) = &~ e^{\F+\bar{\F}}
\left[ \cz(\F) +\bar{\cz}(\bar{\F}) \right]~,\cr
P(\F) = & -  e^{3\F} V(\cz(\F)) \cr}
\ee 

The truly K\"ahler potential $K(\F,\bar{\F})$ is given by \cite{sspace}
\be \lb{kaehler}
 K = -3\ln(-\fracmm{\O}{3})\quad {\rm or} \quad \O=-3e^{-K/3}~, 
\ee
because of the invariance of the action (\ref{stand}) under the supersymmetric
K\"ahler-Weyl transformations
\be \lb{swk}
K(\F,\bar{\F})\to  K(\F,\bar{\F}) +\L(\F) + \bar{\L}(\bar{\F})~, \quad
\ce \to e^{\L(\F)}\ce~, \quad P(\F) \to -  e^{-\L(\F)} P(\F)
\ee 
with an arbitrary chiral superfield parameter $\L(\F)$.
 
The scalar potential (in components) is given by the standard formula 
\cite{crem}
\be \lb{crem}
 \cv (\f,\bar{\f}) =\left. e^{\O} \left\{ \abs{\frac{\pa P}{\pa\F}
+\frac{\pa\O}{\pa\F}P}^2-3\abs{P}^2\right\} \right|~, \ee
where all superfields are restricted to their leading field components,
$\left.\F\right|=\f(x)$.
Equation (\ref{crem}) can be simplified by making use of the K\"ahler-Weyl 
invariance (\ref{swk}) that allows us to choose the gauge
\be  \lb{setone}
P=1 \ee
It is equivalent to the well known fact that the scalar potential (\ref{crem})
is actually governed by the single (K\"ahler-Weyl invariant) potential 
\cite{sspace}
\be\lb{spo}
G(\F,\bar{\F}) = \O +\ln P +\ln \bar{P} \ee
In our case  (\ref{spots}) we have
\be \lb{pot1}
G=e^{\F+\bar{\F}}\left[ \cz(\F) +\bar{\cz}(\bar{\F}) \right]
+ 3(\F+\bar{\F}) + \ln(-V(\cz(\F))+ \ln(-\bar{V}(\bar{\cz}(\bar{\F}))
\ee
Let's now specify our gauge (\ref{ch2}) by choosing the condition
\be \lb{fix}  3\F + \ln(-V(\cz(\F))=0\quad {\rm or}\quad
 V(\cz(\F))=-e^{-3\F}
\ee
that is equivalent to eq.~(\ref{setone}). Then the potential (\ref{pot1}) gets
 simplified to
\be \lb{pot2}
G=\O= e^{\F+\bar{\F}}\left[ \cz(\F) +\bar{\cz}(\bar{\F}) \right]
\ee

Equations (\ref{lt1}), (\ref{lt2}) and (\ref{pot2}) are the simple one-to-one 
algebraic relations between a holomorphic function $F(\car)$ in our modified 
supergravity action (\ref{action}) and a holomorphic function $\cz(\F)$ 
entering the potential (\ref{pot2}) and defining the scalar potential 
(\ref{crem}) as
\be \lb{cpot}
\cv =\left. e^G \left[ \left(\fracmm{\pa^2 G}{\pa\F\pa{\bar{\F}}}\right)^{-1}
\fracmm{\pa G}{\pa\F}\fracmm{\pa G}{\pa\bar{\F}} -3\right] \right|
\ee 
in the classically equivalent  scalar-tensor supergravity. The latter 
can used for embedding the standard slow-roll inflation \cite{inf} into 
supergravity. In our setup that correspondence can be promoted further, by 
embedding the slow-roll inflation into the `purely geometrical' 
higher-derivative supergravity theory (\ref{action}), in terms of a single 
holomorphic function. For our purposes below, we can restrict our attention to
eqs.~(\ref{pot2}) and (\ref{cpot}) in terms of a function $\cz(\F)$ only. 

\section{No-scale modified supergravity}

The no-scale supergravity arises by demanding the scalar potential (\ref{cpot})
to vanish. It results in the vanishing cosmological constant without 
fine-tuning \cite{noscale}. The no-scale  supergravity potential $G$ has to 
obey the non-linear 2nd-order partial differential equation
\be \lb{nleq}
3\fracmm{\pa^2 G}{\pa\F\pa{\bar{\F}}}
=\fracmm{\pa G}{\pa\F}\fracmm{\pa G}{\pa\bar{\F}}
\ee 
A gravitino mass $m_{3/2}$ is given by the vacuum expectation value 
\cite{sspace}
\be \lb{ssb}
m_{3/2}= \VEV{e^{G/2}} \ee
so that the spontaneous supersymmetry breaking scale can be chosen at will.
 
The well known exact solution to eq.~(\ref{nleq}) is given by
\be \lb{wn}
G = -3\ln(\F +\bar{\F}) \ee
In the recent literature, the no-scale solution (\ref{wn}) is usually 
modified by other terms, in order to achieve the universe with a positive 
cosmological constant --- see e.g., the KKLT mechanism \cite{kklt}.

To appreciate the difference between the standard no-scale supergravity 
solution and our modified supergravity, it is worth noticing that the ansatz 
 (\ref{wn}) is inconsistent with our potential (\ref{pot2}) by any choice
of the function $\cz$. Instead, in our case, demanding eq.~(\ref{nleq}) gives 
rise to the 1st-order non-linear partial differential equation
\be \lb{first}
3\left( e^{\bar{\F}}X' + e^{\F}\bar{X}'\right) =\abs{e^{\bar{\F}}X' +
e^{\F}\bar{X}}^2~, 
\ee
where we have introduced the notation
\be \lb{change}
\cz(\F) = e^{-\F}X(\F)~,\qquad X'=\fracmm{dX}{d\F}~, \ee
in order to get the differential equation in its most symmetric and concise 
form. 

Accordingly, the gravitino mass (\ref{ssb}) is given by 
\be \lb{ssb2}
m_{3/2} = \VEV{\exp \ha\left( e^{\bar{\F}}X + e^{\F}\bar{X}\right)}
\ee

I am not aware of any non-trivial holomorphic exact solution to 
eq.~(\ref{first}). Should it obey a holomorphic differential equation of the 
form
\be \lb{holde} 
X'= e^{\F}g(X,\F) 
\ee
with a holomorphic function $g(X,\F)$,  eq.~(\ref{first}) gives rise to a
functional equation,
\be \lb{funce}
3\left( g + \bar{g}\right)= \abs{ e^{\bar{\F}}g +\bar{X}}^2
\ee 

When being restricted to the real variables $\F=\bar{\F}\equiv y$ and 
$X=\bar{X}\equiv x$, eq.~(\ref{first}) reads
\be \lb{real}
6x'=e^y(x'+x)^2~,\quad {\rm where}\quad x'=\fracmm{dx}{dy}
\ee
This equation can be integrated after a change of variables,~\footnote{I am
grateful to A. Starobinsky who pointed it out to me.}
 \be \lb{chvar} 
 x=e^{-y}u~,
\ee
which leads to a quadratic equation with respect to $ u'=du/dy$,
\be \lb{quade}
 (u')^2 - 6u' +6u=0
\ee
It follows
\be
y=\int^u \fracmm{d\x}{3\pm \sqrt{3(3-2\x)}}=\mp \, \sqrt{1-\frac{2}{3}u} 
+\ln\left(\sqrt{3(3-2u)}\pm 3\right) + C~.
\ee

\section{Discussion}

Our main new result is given by eq.~(\ref{pot2}) together with the origin of 
a function $\cz$ out of the action (\ref{action}). It implies the existence 
of new no-scale supergravities based on exact solutions to eq.~(\ref{first}),
with the vanishing cosmological constant and spontaneously broken 
supersymmetry, without fine-tuning.

A generic modified supergravity potential (\ref{pot2}) in terms of the
holomorphic pre-potential $X(\F)$ defined by eq.~(\ref{change}),
\be \lb{simi}
G=\O= e^{\bar{\F}}X(\F) + e^{\F}\bar{X}(\bar{\F}) ~,
\ee  
has essentially the same structure as the supergravity potential of N=2 
extended supergravity (coupled to N=2 vector matter) that is also governed by 
a holomorphic pre-potential. I do not know whether it is merely a coincidence 
or not.   

As regards the non-vanishing (inflaton) scalar potentials from our modified 
supergravity, they have to satisfy the slow-roll conditions 
({\sf cf.} ref.~\cite{sas})
\be
\abs{\fracmm{\pa\cv}{\pa\f}}^2 \ll \fracmm{\pa^2 K}{\pa\f\pa{\bar{\f}}}~\cv^2
\ee
and
\be
\abs{\fracmm{\pa^2\cv}{\pa\f\pa\bar{\f}}
+\fracmm{\pa K}{\pa\bar{\f}}\fracmm{\pa\cv}{\pa\f}}^2 \ll
\left(\fracmm{\pa^2 K}{\pa\F\pa{\bar{\F}}}\right)^2 \cv^2~,
\ee
which guarantee that the potential is sufficiently flat (i.e. suitable for
inflation).

Sometimes it can be achieved by demanding a shift symmetry of the K\"ahler 
potential (i.e. its flatness in one direction), and then perturbing the
 potential around the flat direction --- see e.g., ref.~\cite{shift}. Our 
Ansatz (\ref{pot2}) for the supergravity potential can allow a shift symmetry, 
though it appears to be more restrictive.

In the context of superstring theory, the effective supergravity potential
(\ref{pot2}) may capture some `stringy' features, such as duality and 
maximal curvature --- see e.g., ref.~\cite{gk}.


\newpage

\begin{thebibliography}{99}
\bibitem{pert} R.H. Brandenberger, Lecture Notes Phys. {\bf 646} (2004) 127,
 hep-th/0306071
\bibitem{kall} R. Kallosh,  Lecture Notes Phys. {\bf 738} (2008) 119, 
hep-th/0702059
\bibitem{iihk} M. Iihoshi and S.V. Ketov, Advances in High Energy Physics
(2008) 521389, arXiv:0707.3359 [hep-th]
\bibitem{fluxes} M. R. Douglas and S. Kachru, Rev. Mod. Phys. {\bf 79} (2007) 
733, hep-th/0610102
\bibitem{gket} S. James Gates, Jr. and S. V. Ketov, Superstring-inspired
supergravity as the universal source of inflation and quintessence, 
arXiv:0901.2467[hep-th] 
\bibitem{sot} S. Nojiri and S.D. Odintsov, Int. J. Geom. Meth. Mod. Phys. 
{\bf 4} (2007) 115, hep-th/0601213;\\
T.P. Sotiriou, V. Faraoni, {\it $f(R)$ theories of gravity},
arXiv:0805.1726[hep-th]
\bibitem{gat} S. J. Gates, Jr., Phys. Lett. {\bf B365} (1996) 132 
[hep-th/9508153], and Nucl. Phys. {\bf B485} (1997) 145 [hep-th/9606109]
\bibitem{sspace} S. J. Gates, Jr., M. T. Grisaru, M. Ro\v{c}ek and W. Siegel,
{\it Superspace or 1001 Lessons in Supersymmetry}, Benjamin-Cummings Publ.
 Company, 1983;\\
J. Wess and J. Bagger, Supersymmetry and Supergravity,
Princeton University Press, 1992;\\
I. L. Buchbinder and S. M. Kuzenko, {\it Ideas and Methods of Supersymmetry and
Supergravity}, IOP Publishers, 1998
\bibitem{howe} P. S. Howe and R. W. Tucker, Phys. Lett. {\bf B80} (1978) 138
\bibitem{crem} E. Cremmer, B. Julia, J. Scherk, S. Ferrara, L. Girardello and 
P. van Nieuwenhuizen, Nucl. Phys. {\bf B147} (1979) 105 
\bibitem{inf} A. R. Liddle and D. H. Lyth, {\it Cosmological Inflation and 
Large-Scale Structure}, Cambridge University Press, 2000
\bibitem{noscale} E. Cremmer, S. Ferrara, C. Kounnas and D. V. Nanopoulos, 
Phys. Lett. {\bf B133} (1983) 61
\bibitem{kklt} S. Kachru, R. Kallosh, A. Linde and S. P. Trivedi, Phys. Rev.
{\bf D68} (2003) 046005, hep-th/0301240
\bibitem{sas} M. Sasaki and E. D. Stewart, Progr. Theor. Phys. {\bf 95} (1996)
71, astro-ph/9507001 
\bibitem{shift} M. K. Gaillard, H. Murayama and K. A. Olive, Phys. Lett.
{\bf B355} (1995) 71, hep-ph/9504307;\\
M. Kawasaki, M. Yamaguchi and T. Yanagida, Phys. Rev. Lett. {\bf 85} (2000) 
3572, hep-ph/0004243 
\bibitem{gk} S. J. Gates, Jr. and S. V. Ketov, Class. and Quantum Grav. 
{\bf 17} (2001) 3561, hep-th/0104223.

\end{thebibliography}
\end{document}